\documentclass[apl,amsmath,superscriptaddress,reprint]{revtex4-1}
\usepackage{graphicx}
\usepackage{color}
\usepackage{epstopdf}
\usepackage{xcolor}

\begin{document}

\title{Spin dynamics of an individual Cr atom in a semiconductor quantum dot under optical excitation}

\author{A. Lafuente-Sampietro}
\affiliation{Universit\'{e} Grenoble Alpes, Institut N\'{e}el, F-38000 Grenoble, France}
\affiliation{CNRS, Institut N\'{e}el, F-38000 Grenoble, France}
\affiliation{Institute of Materials Science, University of Tsukuba, Japan}

\author{H. Utsumi}
\affiliation{Institute of Materials Science, University of Tsukuba, Japan}

\author{H. Boukari}
\affiliation{Universit\'{e} Grenoble Alpes, Institut N\'{e}el, F-38000 Grenoble, France}
\affiliation{CNRS, Institut N\'{e}el, F-38000 Grenoble, France}

\author{S. Kuroda}
\affiliation{Institute of Materials Science, University of Tsukuba, Japan}

\author{L. Besombes}\email{lucien.besombes@grenoble.cnrs.fr}
\affiliation{Universit\'{e} Grenoble Alpes, Institut N\'{e}el, F-38000 Grenoble, France}
\affiliation{CNRS, Institut N\'{e}el, F-38000 Grenoble, France}


\begin{abstract}

We studied the spin dynamics of a Cr atom incorporated in a II-VI semiconductor quantum dot using photon correlation techniques. We used recently developed singly Cr-doped CdTe/ZnTe quantum dots (A. Lafuente-Sampietro {\it et al.}, \cite{Lafuente2016}) to access the spin of an individual magnetic atom. Auto-correlation of the photons emitted by the quantum dot under continuous wave optical excitation reveals fluctuations of the localized spin with a timescale in the 10 ns range. Cross-correlation gives quantitative transfer time between Cr spin states. A calculation of the time dependence of the spin levels population in Cr-doped quantum dots shows that the observed spin dynamics is controlled by the exciton-Cr interaction. These measurements also provide a lower bound in the 20 ns range for the intrinsic Cr spin relaxation time.

\end{abstract}

\maketitle


Diluted magnetic semiconductor systems combining high-quality nano-structures and the magnetic properties of transition metal elements are good candidates for the development of single spin
nano-electronic devices \cite{Koenraad2011}. Thanks to their expected long coherence time, localized spin of individual magnetic atoms in a semiconductor host are an interesting media for storing quantum information in the solid state. Optical probing and control of the spin of individual or pairs of magnetic atoms have been obtained in both II-VI \cite{Besombes2004,Goryca2009,LeGall2011,Besombes2012,Besombes2008} and III-V \cite{Kudelski2007,Krebs2013} semiconductors. The variety of magnetic transition elements that could be incorporated in semiconductors gives a large choice of electronic and nuclear spins as well as orbital momentum \cite{Kobak2014,Smolenski2016}. In this context, growth and optical addressing of II-VI semiconductor quantum dots (QDs) containing an individual Cr atom were achieved recently \cite{Lafuente2016}.

Cr is incorporated in II-VI compounds as Cr$^{2+}$ ions carrying a localized electronic spin S = 2 and an orbital momentum L = 2 \cite{Vallin1974}. In addition, most of the Cr isotopes have no nuclear spin. In the presence of a large bi-axial strain, the ground state of the Cr is an orbital singlet with spin degeneracy 2S+1=5. The nonzero orbital momentum of Cr and spin-orbit coupling result in a large sensitivity of its spin to local strain. This large spin to strain coupling, at least two orders of magnitude larger than for magnetic elements without orbital momentum (NV centers in diamond \cite{Barfuss2015,Pigeau2015}, Mn atoms in II-VI semiconductors \cite{Lafuente2015}) makes Cr a very promising spin $qubit$ for the realization of hybrid spin-mechanical systems in which the motion of a microscopic mechanical oscillator would be coherently coupled to the spin state of a single atom \cite{Rabl2010}. Large spin to strain coupling also enhances the spin-phonon interaction ultimately responsible for the spin relaxation and decoherence of an isolated magnetic atom in a semiconductor matrix. A too large interaction with phonons could limit the practical use of Cr as a $qubit$ and an investigation of the spin dynamics in Cr-doped QDs is required.

We show here how we can use the statistics of the photons emitted by a Cr-doped QD to probe the dynamics of the magnetic atom under continuous wave (CW) optical excitation. We performed auto-correlation of the photoluminescence (PL) intensity emitted by individual lines of an isolated Cr-doped QD using a Hanbury Brown and Twiss (HBT) setup. In these start-stop experiments, the detection of the first photon indicates by its energy and polarization that the Cr spin has a given orientation. The probability of detection of a second photon with the same energy and polarization is proportional to the probability of conserving this spin state. The time evolution of this intensity correlation signal is a probe of the spin dynamics in the Cr-doped QD. The auto-correlation signal presents a large bunching revealing a PL intermittency which results from fluctuations of the Cr spin with a timescale in the 10 ns range. Correlation of the intensity emitted by two different lines of the exciton-Cr (X-Cr) complex (namely, cross-correlation), presents an anti-bunching at short delays. A calculation of the time dependence of the spin levels population in Cr-doped QDs shows that the observed spin dynamics in auto-correlation and cross-correlation measurements is governed by the exciton/Cr interaction. These measurements also provide a lower bound in the 20 ns range for the intrinsic Cr spin relaxation time.


To optically probe an individual magnetic atom, Cr are randomly introduced in CdTe/ZnTe self-assembled QDs. QDs are grown by molecular beam epitaxy on ZnTe (001) substrates following the procedure described in Ref. \cite{Wojnar2011}. The amount of Cr is adjusted to allow for the detection of QDs containing 1 or a few Cr atoms. The emission of individual QDs is studied by optical micro-spectroscopy. Permanent magnets can be used to apply a weak magnetic field in both Faraday or Voight geometry. The PL is quasi-resonantly excited with a tunable CW dye laser or by picosecond pulses from a doubled optical parametric oscillator. High refractive index hemispherical solid immersion lens is used to enhance the collection of the single-dot emission. The statistics of the photons emitted by Cr-doped QDs is analyzed using a HBT setup for photon-correlation measurements with a time resolution of about 700 ps \cite{Sallen2011}. The circularly polarized collected light is spectrally dispersed by a 1 m double monochromator before being detected in the HBT setup or by a fast avalanche photodiode (time resolution of 50 ps) for time-resolved measurements. Under our experimental conditions with count rates of a few kHz, the measured photons pair distribution in the HBT setup yields, after normalization, the second order correlation function of the PL intensity g$^{(2)}(\tau)$.

\begin{figure}[hbt]
\includegraphics[width=3.3in]{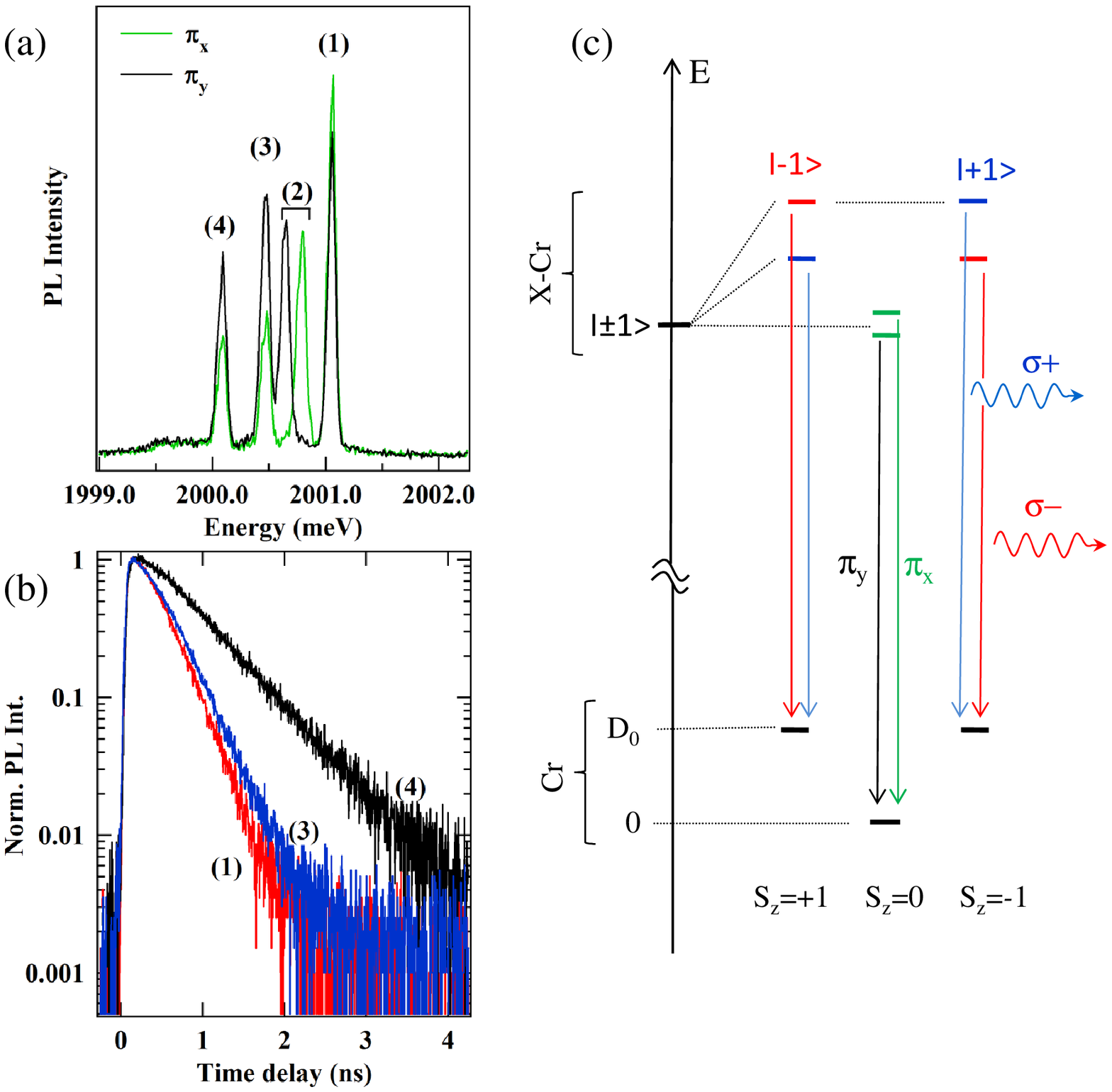}
\caption{(a) Low temperature (T=5K) linearly polarized PL recorded along two orthogonal directions of the exciton in Cr-doped QD (X-Cr). (b) Time resolved PL of the X-Cr lines (1), (3) and (4). (c) Scheme of the energy levels in a Cr-doped QD showing the magnetic ground states S$_z$=0 and S$_z$=$\pm$1 of the Cr and the corresponding bright exciton states ($|\pm1\rangle$) coupled to the spin of the Cr (X-Cr). The higher energy Cr states S$_z$=$\pm$2, unpopulated at T=5K, are not displayed.}
\label{Fig1}
\end{figure}

The PL of X-Cr in a Cr-doped QD is reported in Fig.~\ref{Fig1}(a). Because of the large magnetic anisotropy of the Cr spin, $D_0S_z^2$, induced by the local bi-axial strain in self-assembled QDs, the Cr spin thermalizes to its magnetic ground states $S_z$=0 and $S_z$=$\pm$1. The exchange interaction with the spin of the confined bright exciton $|\pm1\rangle$ splits the Cr spin states $S_z$=$\pm$1 and three main lines (labeled (1), (2) and (3)) are observed in the PL spectra \cite{Lafuente2016}. Most of the dots also present a low symmetry (lower than C$_{2v}$) and the central line associated with $S_z$=0 (line (2)), is split and linearly polarized along two orthogonal directions, by (i) the long-range e-h exchange interaction in a QD with an in-plane shape anisotropy and/or (ii) the short range e-h exchange interaction in the presence of valence band mixing. An additional line (labeled (4)) also appears on the low-energy side of the PL spectra. As presented in Fig.~\ref{Fig1}(b), this line presents a longer lifetime. It arises from a dark exciton ($|\pm2\rangle$) which acquires some oscillator strength by a mixing with a bright exciton interacting with the same Cr spin state \cite{Lafuente2016}. This dark/bright exciton mixing is induced by the e-h exchange interaction in a confining potential of low symmetry \cite{Zielinski2015}.


\begin{figure}[hbt]
\includegraphics[width=3.3in]{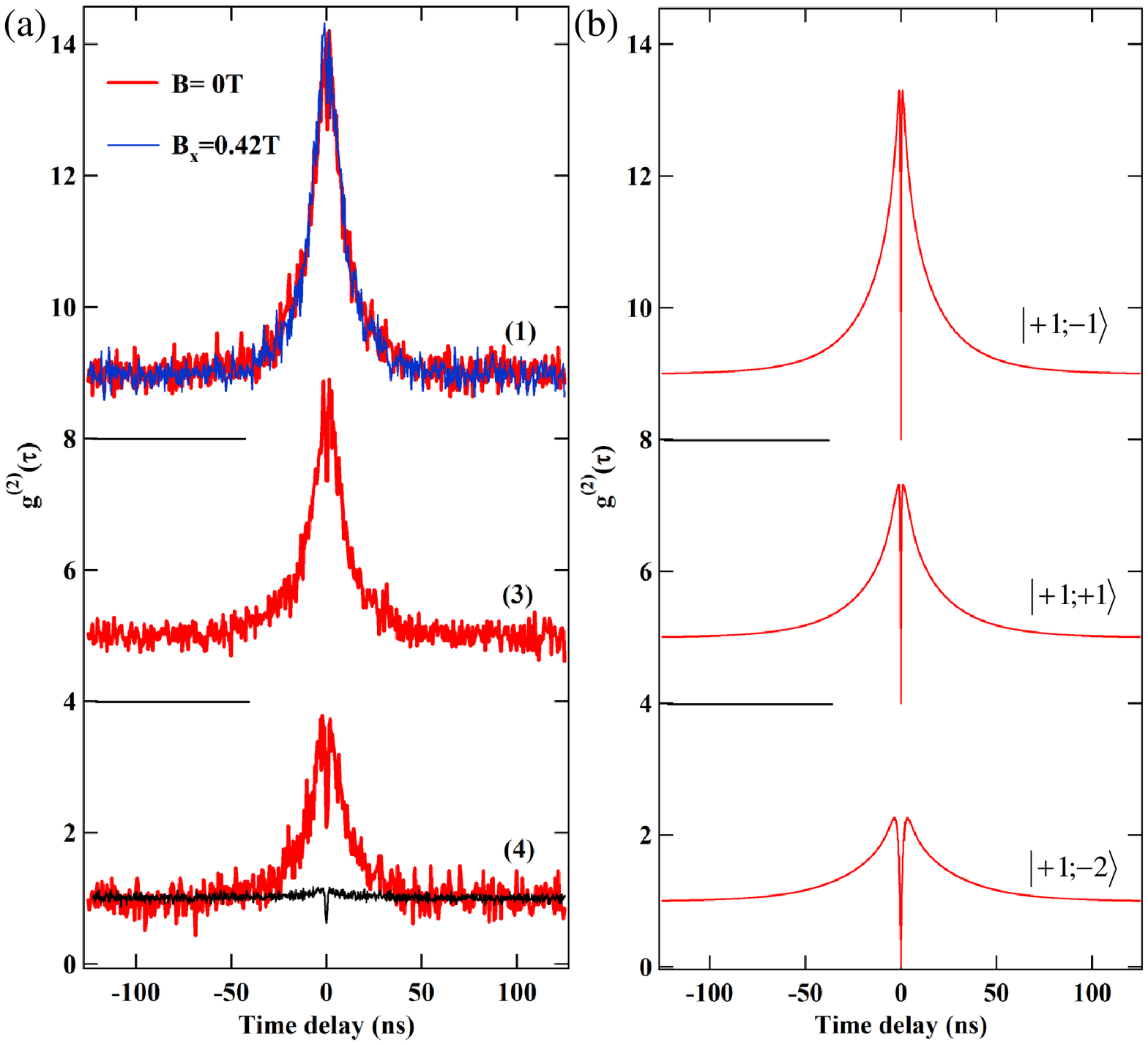}
\caption{(a) Auto-correlation of the PL intensity collected in circular polarization on the X-Cr lines (1), (3) and (4) (see Fig.~\ref{Fig1}) and compared with the auto-correlation of the exciton in a non-magnetic QD (black line). The curves are shifted for clarity. For line (1), the auto-correlation is also recorded under a transverse magnetic field (blue line). (b) Corresponding calculated auto-correlation for three X-Cr states $|S_z,X\rangle$ with $\tau_g$=4ns, $\tau_{Cr}$=50ns, $T_{eff}$=30K, $\tau_b$=0.25ns, $\tau_d$=5ns and the Cr-doped QD parameters of Ref.\cite{Lafuente2016}.}
\label{Fig2}
\end{figure}

To observe the time fluctuations of the Cr spin under CW optical excitation, we used the statistics of time arrivals of the photons emitted by a Cr-doped QD given by the second order correlation function g$^{(2)}(\tau)$ of the PL intensity. Fig.~\ref{Fig2}(a) shows g$^{(2)}(\tau)$ for the lines (1), (3) and (4) recorded in circular polarization. These signals are compared with the auto-correlation obtained for the PL of a non-magnetic QD which is characteristic of a single-photon emitter with a deep (anti-bunching) at short delays. The width of this anti-bunching is given by the lifetime of the emitter and the generation rate of excitons and its depth is limited by the time resolution of the HBT setup. A similar experiment performed on X-Cr still presents a reduced coincidence rate near zero delay, but it is mainly characterized by a large photon bunching with a full width at half maximum (FWHM) in the 20 ns range. The amplitude of the bunching can reach 6 for line (1) and is slightly weaker for the lower energy lines. This bunching reflects an intermittency in the emission of a given line of the QD coming from fluctuations of the Cr spin in a 10 ns timescale. The photon bunching is not affected by a weak transverse magnetic field ($B_x=0.42$T in Fig.~\ref{Fig1}(a)). This confirms the presence of a large strain induced magnetic anisotropy $D_0S_z^2$ which splits the Cr and X-Cr spin states and blocks their precession in an external magnetic field.


To identify the main contribution to the observed spin fluctuations, we modelled the auto-correlation of the PL of X-Cr using the full spin level structure of a Cr-doped QD. We calculated the time evolution of the population of the twenty X-Cr states in the excited state of the QD and five Cr states in the ground state by solving numerically the master equation for the corresponding 25 x 25 density matrix $\rho$. The time evolution of the density matrix including relaxation and dephasing processes in the Lindblad form is given by $\partial \rho/\partial t=-i/\hbar[{\cal H},\rho]+L\rho$ where ${\cal H}$ is the Hamiltonian of the complete system ($X$-Cr and Cr) and $L\rho$ describes the coupling or decay channels resulting from an interaction with the environment \cite{Exter2009,Roy2011}. The energy levels of the Cr are controlled by the magnetic anisotropy $D_0S_z^2$. The X-Cr Hamiltonian, described in details in Ref.\cite{Lafuente2016}, contains the energy of the Cr spin states, the carriers-Cr exchange interactions, the electron-hole exchange interaction in a confining potential of low symmetry and the structure of the valence band including heavy-hole/light-hole mixing. $D_0$ in the Cr Hamiltonian and the parameters in the X-Cr Hamiltonian cannot be precisely extracted from the zero magnetic field PL (Fig.~\ref{Fig1}(a)). For a qualitative description of the observed spin dynamics, we use in the model typical Cr-doped QD parameters extracted from magneto-optics measurements presented in Ref. \cite{Lafuente2016}. These parameters give a X-Cr splitting and a dark/bright excitons mixing similar to the one observed in the QD discussed in this article.

\begin{figure}[hbt]
\includegraphics[width=3.3in]{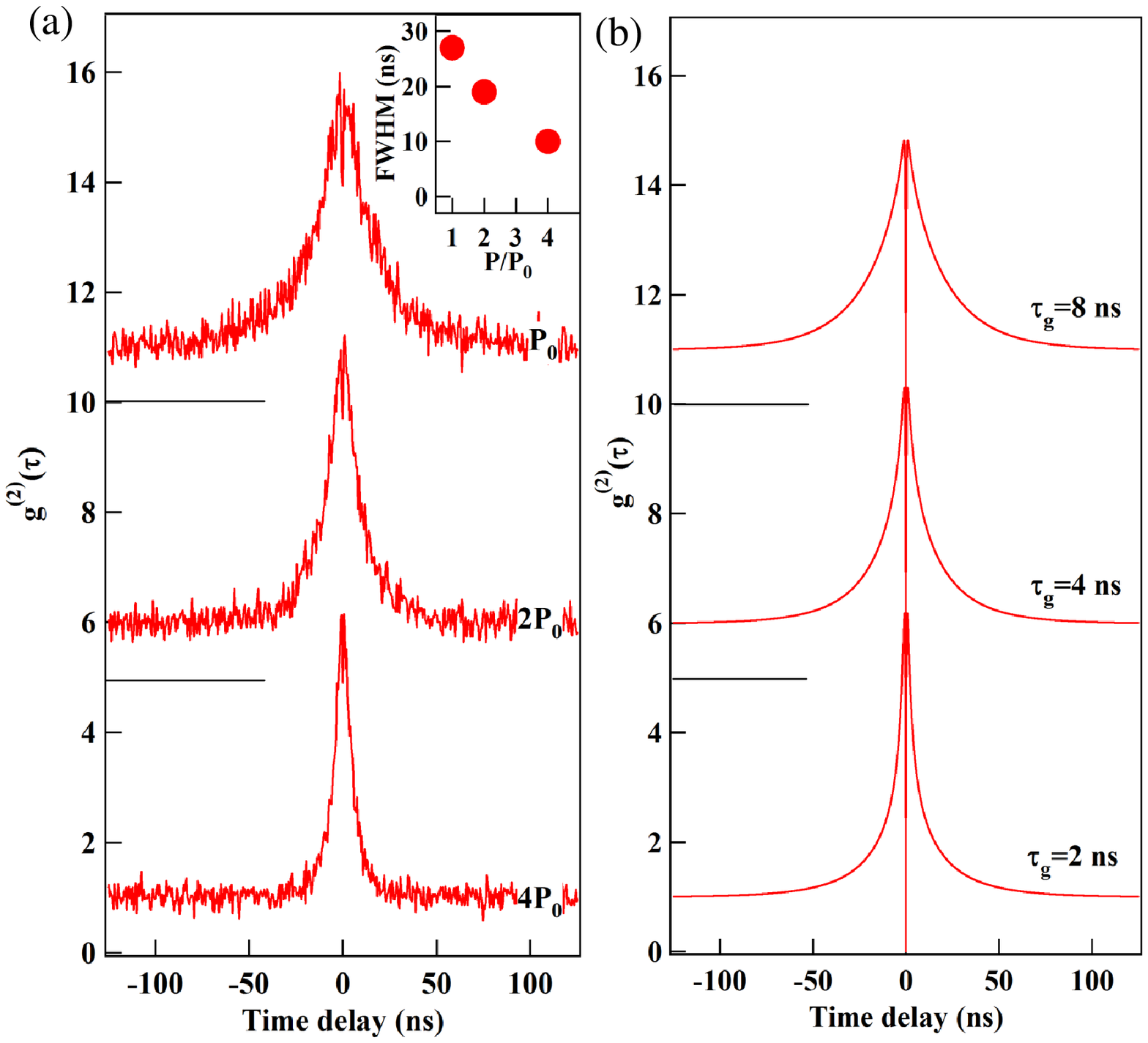}
\caption{(a) Auto-correlation of the PL intensity recorded in circular polarization on the high energy X-Cr line (1) for different excitation powers. The inset shows the corresponding FWHM of the bunching signal versus excitation power. (b) Calculated excitation power dependence of the auto-correlation of the X-Cr state $|+1,-1\rangle$.}
\label{Fig3}
\end{figure}

We took into account a spin relaxation of the Cr (alone or in the exchange field of the exciton), $\tau_{Cr}$, describing relaxation channels involving a change of the Cr spin by one unit. The transition rates $\Gamma_{\gamma\rightarrow\gamma'}$ between the different states of the Cr or X-Cr depend on their energy separation $E_{\gamma\gamma'}=E_{\gamma'}-E_{\gamma}$. Here we use $\Gamma_{\gamma\rightarrow\gamma'}$=1/$\tau_{Cr}$ if $E_{\gamma\gamma'}<0$ and $\Gamma_{\gamma\rightarrow\gamma'}$=1/$\tau_{Cr}e^{-E_{\gamma\gamma'}/k_BT_{eff}}$ if $E_{\gamma\gamma'}>0$ \cite{Govorov2005,Cao2011}. This accounts for a thermalization among the 5 Cr levels on one side and the 20 X-Cr levels on the other side with an effective spin temperature $T_{eff}$. The optical excitation ($\tau_g=1/\Gamma_g$), the exciton recombination ($\tau_b$ (bright) and $\tau_d$ (dark)) and the Cr spin relaxation ($\tau_{Cr}$) producing an irreversible population transfer from level $j$ to level $i$ are described by Lindblad terms $L_{j\rightarrow i}\rho=1/(2\tau_{j\rightarrow
i})(2|i\rangle\langle j|\rho|j\rangle\langle i|-\rho|j\rangle\langle j|-|j\rangle\langle j|\rho)$ where $\tau_{j\rightarrow i}$ is the relaxation time from level $j$ to level $i$ \cite{Jamet2013}.

We considered that the CW optical excitation creates bright and dark excitons with a generation rate $\Gamma_g/4$. They recombine with a rate $1/\tau_b$ and $1/\tau_d$ respectively. The time evolution of $\rho_{|S_z,X\rangle}(t)/\rho_{|S_z,X\rangle}(\infty)$ with the initial condition $\rho_{|S_z\rangle}(0)=1$ (Cr spin state $S_z$ after the recombination of the exciton) accounts for the auto-correlation of the emission of the X-Cr state $|S_z,X\rangle$.

The results obtained with a Cr spin-flip time $\tau_{Cr}$=50 ns and a generation time $\tau_g=4 ns$ are presented in Fig.~\ref{Fig2}(b) for the X-Cr states $|+1,-1\rangle$ (high energy bright exciton), $|+1,+1\rangle$ (low energy bright exciton) and $|+1,-2\rangle$ (dark exciton). The width of the calculated bunching signals and the evolution of their amplitude from the high to low energy line are in qualitative agreement with the measured dynamics.


\begin{figure}[hbt]
\includegraphics[width=3.3in]{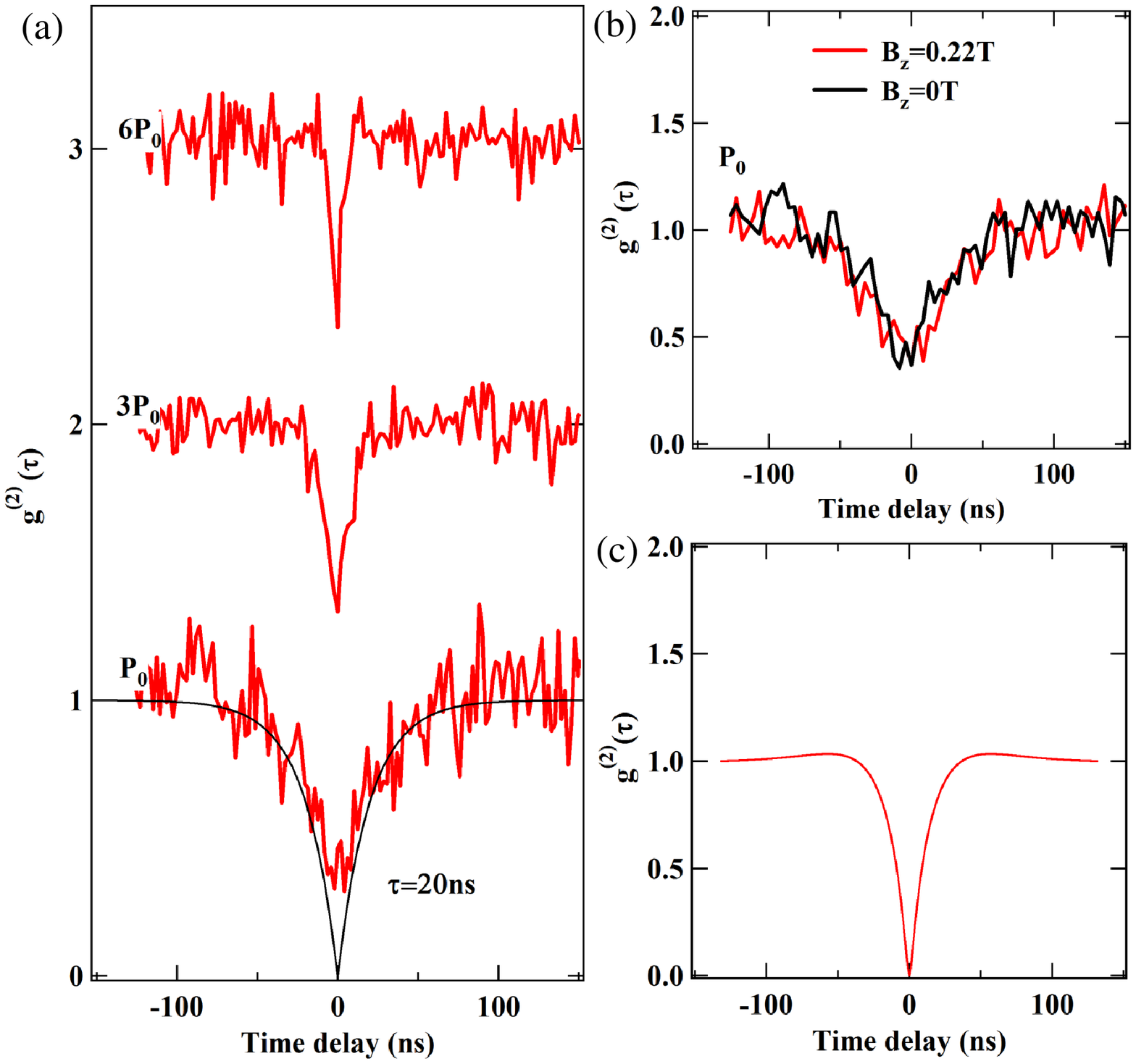}
\caption{(a) Correlation signal of the PL intensity of lines (1) and (3) recorded in the same circular polarization (cross-correlation) for three different excitation powers. The curves are shifted for clarity. The black line is an exponential fit with a characteristic time $\tau$=20ns. (b) Longitudinal magnetic field dependence of the cross-correlation signal obtained at low excitation power. (c) Calculated cross-correlation signal ($\rho_{|S_z=+1,+1\rangle}(t)/\rho_{|S_z=+1,+1\rangle}(\infty)$ with $\rho_{|S_z=-1\rangle}(0)=1$) with $\tau_g$=8ns and the parameters used in the modelling presented in Fig.~\ref{Fig2}.}
\label{Fig4}
\end{figure}

However, one should note that the observed spin dynamics depends on the optical excitation power. Increasing the excitation power significantly reduces the width of the bunching (Fig.~\ref{Fig3}(a)). This increase of the Cr spin fluctuations under optical excitation is well described by the proposed model. This is illustrated in Fig.~\ref{Fig3}(b) which presents the calculated auto-correlation of the X-Cr state $|+1,-1\rangle$ (high energy line) for a fixed Cr spin relaxation time $\tau_{Cr}$=50 ns and three different generation rates. An increase of the generation rate reproduces the observed reduction of the auto-correlation width. Within the X-Cr complex, the electron-Cr exchange interaction and the hole-Cr exchange interaction in the presence of heavy-hole/light-hole mixing can both induce spin-flips of the Cr. Though weak, the probability of such spin flips increases with the occupation of the QD with an exciton and dominates the spin dynamics in the high excitation regime required for the photon correlation measurements.

This excitation power dependence shows that the measured width of the bunching is not limited by the intrinsic Cr spin relaxation time $\tau_{Cr}$. This gives a lower bound for the intrinsic Cr spin relaxation in the 20 ns range. A shorter value would impose, at low excitation intensity, faster spin fluctuations than observed experimentally. The Cr spin relaxation time is ultimately controlled by the interaction with acoustic phonons and could depend on the optical excitation through the generation of non-equilibrium acoustic phonons during the relaxation of injected carriers \cite{Hundt2005,Kneip2006}. It is however expected to be much longer than the observed dynamics \cite{Cao2011} and cannot be determined with these measurements which require a large photon count rate.


To analyze more in detail the X-Cr spin relaxation channels, cross-correlation measurements were performed on the PL emitted by the high energy and the low energy lines in the same circular polarization. The cross-correlation shows a large anti-bunching with a FWHM in the 10 ns range and g$^{(2)}$(0)$\approx$0.3 (Fig.~\ref{Fig4}(a)). Whereas the auto-correlation probes the time dependence of the probability for the spin of the Cr to be conserved, this cross-correlation is a probe of the spin transfer time between the spin states S$_z$=+1 and S$_z$=-1. As for the auto-correlation, the cross-correlation strongly depends on the excitation power. At weak excitation, a spin transfer time of about 20 ns is observed. It is accelerated with the increase of the excitation power (Fig.~\ref{Fig4}(a)). This transfer time could be controlled by anisotropic in-plane strain which couples Cr spin states separated by two units through an additional term in the Cr fine structure Hamiltonian of the form $E(S_x^2-S_y^2)$ \cite{Lafuente2016}. However, even at low excitation power, the measured transfer time is not affected by a longitudinal magnetic field (Fig.~\ref{Fig4}(b)). This shows that for such QD the strain anisotropy term $E$ is weak and is not the main parameter controlling the transfer time between the spin states S$_z$=$\pm$1. Indeed, the anti-bunching in the cross-correlation signal can also be qualitatively reproduced by the proposed density matrix model model with the parameters used for the simulation of the auto-correlation (Fig.~\ref{Fig4}(c)). The spin transfer time is dominated by spin-flips induced by the exciton/Cr interaction.


To conclude, we have shown that we can use the statistics of the photons emitted by a Cr-doped QD to probe the spin dynamics of the magnetic atom under CW optical excitation. The measured spin fluctuation time, in the 10 ns range, significantly depends on the excitation power. A calculation of the spin states populations using a density matrix formalism shows that under optical excitation the spin dynamics in Cr-doped QDs is dominated by carrier-Cr coupling. Nevertheless, these experiments provide a lower bound for the intrinsic Cr spin relaxation time in the 20 ns range. Directly probing this intrinsic Cr spin relaxation time, ultimately controlled by the spin-lattice coupling, would require measuring the dynamics in the absence of injected carriers and/or non-equilibrium phonons.

\begin{acknowledgements}

The authors acknowledge financial support from the Labex LANEF for the Grenoble-Tsukuba collaboration. The study in Tsukuba has partially been supported by Grant-in-Aid for Scientific Research on Innovative Areas and Exploratory Research. The work in Grenoble was realized in the framework of the Commissariat \`{a}  l'Energie Atomique et aux Energies Alternatives (Institut Nanosciences et Cryog\'{e}nie) / Centre National de la Recherche Scientifique (Institut N\'{e}el) joint research team NanoPhysique et Semi-Conducteurs.

\end{acknowledgements}

\end{document}